\documentclass{article}%
\usepackage{amsfonts}
\usepackage{amsmath}
\usepackage{amssymb}
\usepackage{graphicx}%
\providecommand{\U}[1]{\protect\rule{.1in}{.1in}}

\begin{document}

\title{Entropic Dynamics: Quantum Mechanics from Entropy and Information Geometry}
\author{Ariel Caticha\\{\small Department of Physics, University at Albany--SUNY, Albany, NY 12222,
USA}}
\date{}
\maketitle

\begin{abstract}
Entropic Dynamics (ED) is a framework in which Quantum Mechanics (QM) is
derived as an application of entropic methods of inference. The magnitude of
the wave function is manifestly epistemic: its square is a probability
distribution. The epistemic nature of the phase of the wave function is also
clear: it controls the flow of probability. The dynamics is driven by entropy
subject to constraints that capture the relevant physical information. The
central concern is to identify those constraints and how they are updated.
After reviewing previous work I describe how considerations from information
geometry allow us to derive a phase space geometry that combines Riemannian,
symplectic, and complex structures. The ED that preserves these structures is
QM. The full equivalence between ED and QM is achieved by taking account of
how gauge symmetry and charge quantization are intimately related to quantum
phases and the single-valuedness of wave functions.

\end{abstract}

\section{Introduction}

Ever since its origin in 1925 the conceptual foundations of Quantum Mechanics
(QM) have been a source of controversy. Part of the problem is that quantum
mechanics is often conceived as a generalization of classical mechanics with
an added quantum indeterminism. The linear Hilbert space structure is given
priority while the probabilistic structure is appended almost as an
afterthought. The unfortunate consequence is that in the standard approach to
QM the dynamical and the probabilistic aspects of quantum theory are not quite
compatible with each other. Two separate and irreconcilable modes of wave
function evolution are postulated: one is the linear and deterministic
Schr\"{o}dinger evolution and the other is the probabilistic wave function
collapse. The problem is that it is not clear why measurement processes should
be any different from other regular physical processes. Closely related to
these issues is the interpretation of the quantum state itself. Does the wave
function represent the actual state of the system (its \emph{ontic} state) or
does it represent our knowledge about the system (an \emph{epistemic}
state)?\footnote{Excellent reviews with extended references to the literature
are given in \emph{e.g.} \cite{Stapp 1972}-\cite{Leifer 2014}.}

Entropic Dynamics (ED) resolves these problems by placing the probabilistic
aspects of QM at the forefront. The Schr\"{o}dinger equation is derived as an
application of entropic methods of inference.\footnote{The principle of
maximum entropy as a method for inference can be traced to the pioneering work
of E. T. Jaynes \cite{Jaynes 1957}-\cite{Jaynes 2003}. For a pedagogical
overview of Bayesian and entropic inference and further references see
\cite{Caticha 2012}.} When constructing models of this kind the first step is
to specify the subject matter, the ontology. This step is not trivial: are we
talking about particles or about particle detectors? Once the choice is made
the inferences are driven by entropy subject to information expressed by
constraints. The challenge is to identify the right constraints; it is through
them that the \textquotedblleft physics\textquotedblright\ is introduced
\cite{Caticha 2010}-\cite{Carrara Caticha 2017}. This view that the wave
function $\Psi$ is a fully epistemic concept turns out to be extremely
restrictive because it is not sufficient to merely interpret the probability
$|\Psi|^{2}$ as a state of knowledge. It is also required that all changes in
$\Psi$, which include \emph{both} the unitary time evolution \emph{and} the
wave function collapse, be obtained as a consequence of entropic and Bayesian
updating rules.

The literature on the attempts to reconstruct quantum mechanics is vast (see
\emph{e.g.}, \cite{Nelson 1985}-\cite{tHooft 2016} and references therein) and
there are many approaches that are based on information theory (see
\emph{e.g.}, \cite{Wootters 1981}-\cite{Reginatto Hall 2012}). What
distinguishes ED is a strict adherence to Bayesian and entropic methods and
also a central concern with the nature of time. In ED \textquotedblleft
entropic\textquotedblright\ time is a book-keeping device \emph{designed} to
keep track of change. The construction of entropic time involves several
ingredients. One must introduce the notion of an `instant'; one must show that
these instants are suitably ordered; and finally one must define a convenient
measure of the duration or interval between the successive instants. It turns
out that an arrow of time is generated automatically. Entropic time is
intrinsically directional.\ 

This paper contains a review of the ED framework \cite{Caticha 2010}%
-\cite{Carrara Caticha 2017} and extends its formalism in three new
directions. The first new contribution is a rather straightforward variation
on the ED theme. I show that some of the relevant dynamical information ---
which in the past had been imposed through constraints --- can alternatively
be supplied through the prior distribution.

The second new contribution builds on and extends recent work in collaboration
with N. Carrara. The issue addressed in \cite{Carrara Caticha 2017} is whether
ED is fully equivalent to QM or does it merely reproduce a subset of its
solutions. This kind of problem was first pointed out long ago by Takabayasi
\cite{Takabayasi 1952} in the context of the hydrodynamical interpretation of
QM, and later revived by Wallstrom \cite{Wallstrom 1989}\cite{Wallstrom 1994}
as an objection to Nelson's stochastic mechanics. Wallstrom's objection is
that stochastic mechanics leads to phases and wave functions that are either
both multi-valued or both single-valued. Both alternatives are unsatisfactory
because on one hand QM requires single-valued wave functions, while on the
other hand single-valued phases exclude states that are physically relevant
(\emph{e.g.}, states with non-zero angular momentum).

The resolution of Wallstrom's objection in the context of ED involves two
related ingredients.\footnote{A hint towards a satisfactory resolution of
Wallstrom's objection is found in Takabayasi's later work which incorporates
spin into his hydrodynamical approach \cite{Takabayasi 1983}. Although here we
focus on non-spinning particles our choice of constraints can be generalized
to particles with spin $1/2$ --- a project to be addressed in a future
publication.} The first recognizes the physical fact that quantum phases are
intimately related to the local electromagnetic gauge symmetry. One cannot
properly understand one without taking the other into account. The second
ingredient recognizes the experimental fact of charge quantization --- that
charges occur in multiples of some basic unit. The new and welcome bonus is a
deeper understanding of the connection between the probabilistic structure of
QM, its linearity, the quantization of charge, and the single-valuedness of
wave functions.\footnote{The issue of single-valuedness has been discussed
from the different perspective of group representations by Pauli \cite{Pauli
1939}\cite{Merzbacher 1962}.}

The essence of Entropic Dynamics is the entropic updating of probabilities
through information supplied by constraints. The central concern, therefore,
is how these constraints are chosen, and in particular, how the constraints
themselves are updated. One early insight due to Nelson was that several
aspects of quantum behavior could be described as a Brownian motion subject to
a conserved energy: stochastic mechanics is a non-dissipative diffusion
\cite{Nelson 1979}. Fortunately this was an idea that could be directly
imported from Nelson's realistic stochastic mechanics into our epistemic ED.
Although in ED energy is a purely epistemic notion, its conservation provides
an effective criterion for choosing evolving constraints. This leads to a
fully Hamiltonian formalism with its attendant symplectic structure, action
principle, Poisson brackets, and so on. Unfortunately this approach, while
fully satisfactory in a non-relativistic setting, fails in curved space-times
where the concept of a globally conserved energy may not exist.

The third new contribution in this paper is concerned with developing a
geometric criterion for updating constraints that does not rely on the notion
of a conserved energy. The fact that QM has a rich geometrical structure
\cite{Kibble 1979}-\cite{Ashtekar Schilling 1998} including a deep connection
to information geometry \cite{Wootters 1981} has been explored by many
authors. There is, for example, the application to quantum statistical
inference developed by Brody and Hughston \cite{Brody Hughston 1997}, and the
operational framework for the analysis of measurements described in the work
of Mehrafarin \cite{Mehrafarin 2005} and Goyal \cite{Goyal 2010}. The work of
Reginatto and Hall \cite{Reginatto Hall 2011}\cite{Reginatto Hall 2012} is
particularly relevant to us. They impose a symplectic structure in the phase
space $\{\rho,\Phi\}$ of probability distributions $\rho$ and their conjugate
momenta $\Phi$. To motivate the choice of Hamiltonian flow they extend the
natural information metric over the ensemble space $\{\rho\}$ to the whole
phase space $\{\rho,\Phi\}$ in a way that maintains the symplectic structure.

The approach we pursue here differs by virtue of being based on a different
epistemic goal --- to identify constraints for entropic inference. More
specifically we \emph{derive} the symplectic structure rather than impose it
as a starting point. In ED the degrees of freedom are the probability
densities $\rho(x)$ and the phase fields $\Phi(x)$. The latter represent the
constraints that control the flow of probabilities. Our goal too is to extend
the information geometry from the \textquotedblleft ensemble\textquotedblright%
\ space $\{\rho\}$ to the full \textquotedblleft
ensemble-phase\ space\textquotedblright\ $\{\rho,\Phi\}$, which we abbreviate
as \textquotedblleft e-phase\textquotedblright\ space. To do this we impose a
symmetry that is natural in a probabilistic setting: we extend the well-known
spherically\ symmetric information geometry from the ensemble space $\{\rho\}$
to the full e-phase space $\{\rho,\Phi\}$. Then we show that the extended
metric describes quantum geometry, that is, it includes the Riemannian, the
complex, and the symplectic structures that we recognize as quantum geometry.

The desired geometric criterion for updating constraints is a dynamics that
preserves these structures. Therefore ED is a Hamiltonian dynamics. The most
natural Hamiltonian is simply constructed from the extended metric by
recognizing the privileged ontic role of position. The resulting dynamics is
described by the Schr\"{o}dinger equation.

This paper focuses on the derivation of the Schr\"{o}dinger equation but the
ED approach has been applied to a variety of other topics including the
quantum measurement problem \cite{Johnson Caticha 2011}\cite{Vanslette Caticha
2016}; momentum and uncertainty relations \cite{Nawaz Caticha 2011}; the
Bohmian limit \cite{Bartolomeo Caticha 2015}\cite{Bartolomeo Caticha 2016} and
the classical limit \cite{Demme Caticha 2016}; extensions to curved spaces
\cite{Nawaz et al 2015} and to relativistic fields \cite{Ipek Caticha
2014}\cite{Ipek Abedi Caticha 2016} are also available.

\section{The statistical model}

We consider $N$ particles living in a flat Euclidean space $\mathbf{X}$ with
metric $\delta_{ab}$. The first assumption is that the particles have
\emph{definite }positions\emph{ }$x_{n}^{a}$, collectively denoted by\emph{
}$x$\emph{, }and it is their \emph{unknown} values that we wish to infer. (The
index $n$ $=1\ldots N$ labels the particles, and $a=1,2,3$ the three spatial
coordinates.) The configuration space for $N$ particles is $\mathbf{X}%
_{N}=\mathbf{X}\times\ldots\times\mathbf{X}$.

Since the positions are unknown the main target of our attention is the
probability distribution $\rho(x)$. Incidentally, this already addresses that
old question of determinism vs. indeterminism that has troubled quantum
mechanics from the outset. Inference techniques are designed to cope with
insufficient information. In an inference approach to quantum theory one
starts by accepting uncertainty, probability, and indeterminism as the
expected and inevitable norm that requires no explanation; it is the certainty
and determinism of classical mechanics that demand explanations (see
\emph{e.g.}, \cite{Demme Caticha 2016}).

The assumption that the particles have definite positions that happen to be
unknown is a major departure from the standard Copenhagen interpretation
according to which observable quantities do not in general have definite
values. In the standard approach observables can attain definite values but
only as the result of a measurement. In contrast, positions in ED play a very
special role: they define the ontic state of the system. For example, in the
double slit experiment, we might not know which slit the particle goes
through, but the particle definitely goes through one slit or the other. The
wave function, on the other hand, is a purely epistemic notion; it defines our
epistemic state about the system. All other quantities, such as energy or
momentum, are epistemic in that they reflect properties of the wave function,
not properties of the particles \cite{Johnson Caticha 2011}-\cite{Nawaz
Caticha 2011}. These values are not quite \textquotedblleft
created\textquotedblright\ by the measurement, but rather inferred from them.
Such quantities may, therefore, be more aptly referred to as \textquotedblleft
inferables\textquotedblright\ rather than \textquotedblleft
observables\textquotedblright.\footnote{I thank Kevin Vanslette for suggesting
this terminology.}

Once the microstates $x\in\mathbf{X}_{N}$ are identified we proceed to the
dynamics. The goal is to find the probability density $P(x^{\prime}|x)$ of a
step from an initial position $x\in\mathbf{X}_{N}$ to a new neighboring
$x^{\prime}\in\mathbf{X}_{N}$ by maximizing the entropy,
\begin{equation}
\mathcal{S}[P,Q]=-\int dx^{\prime}\,P(x^{\prime}|x)\log\frac{P(x^{\prime}%
|x)}{Q(x^{\prime}|x)}~, \label{entropy a}%
\end{equation}
relative to a prior $Q(x^{\prime}|x)$, and subject to the appropriate
constraints specified below. (For notational simplicity in multidimensional
integrals such as (\ref{entropy a}) we will write $dx^{\prime}$ instead of
$d^{n}x^{\prime}$.)

\paragraph*{The prior--}

In ED one does not explain why motion happens --- this is a \textquotedblleft
mechanics without a mechanism.\textquotedblright\ Instead our task is to
produce an estimate of what kind of motion one might reasonably expect.
\emph{The main dynamical assumption is that the particles follow trajectories
that are continuous.} This introduces an enormous simplification because it
implies that motion can be analyzed as the accumulation of many
infinitesimally short steps and, therefore, our first task is to find the
transition probability $P(x^{\prime}|x)$ for an infinitesimally short step.

The prior $Q(x^{\prime}|x)$ in eq.(\ref{entropy a}) expresses our state of
knowledge \emph{before} any information about the details of a particular
motion are taken into account. Accordingly, we adopt a prior that incorporates
the information that the particles' steps are infinitesimally short, but is
otherwise maximally uninformative. We want a prior that reflects translational
and rotational invariance, and is ignorant about any correlations. Such a
prior can itself be derived from the principle of maximum entropy. Indeed,
maximize
\begin{equation}
S[Q]=-\int dx^{\prime}\,Q(\Delta x)\log\frac{Q(\Delta x)}{\mu(\Delta x)}~,
\end{equation}
where $\Delta x_{n}^{a}=x_{n}^{\prime a}-x_{n}^{a}$, relative to the uniform
measure $\mu(\Delta x)$, subject to normalization, and $N$ independent
constraints that enforce short steps and rotational invariance,
\begin{equation}
\langle\delta_{ab}\Delta x_{n}^{a}\Delta x_{n}^{b}\rangle=\kappa_{n}%
~,\quad(n=1\ldots N)~,
\end{equation}
where $\kappa_{n}$ are small constants. The result is a product of Gaussians,
\begin{equation}
Q(x^{\prime}|x)\propto\exp-\frac{1}{2}%
{\displaystyle\sum\nolimits_{n}}
\alpha_{n}\delta_{ab}\Delta x_{n}^{a}\Delta x_{n}^{b}~, \label{prior}%
\end{equation}
where the $\alpha_{n}$ are Lagrange multipliers that will eventually be taken
to infinity in order to enforce infinitesimally short steps. For now we just
anticipate that the $\alpha_{n}$ are constants that are independent of $x$ but
may depend on the index $n$ in order to describe non-identical particles. Next
we specify the constraints.

\paragraph*{The drift potential constraint--}

The physical information that the motion of the particles can be both
directional and highly correlated --- including such effects as entanglement
--- is introduced through \emph{a single constraint} involving a
\textquotedblleft drift\textquotedblright\ potential $\phi(x)=\phi(x_{1}\ldots
x_{N})$ that is a function in configuration space, $x\in\mathbf{X}_{N}$. We
impose that the displacements $\Delta x_{n}^{a}$ are such that the expected
change of the drift potential $\left\langle \Delta\phi\right\rangle $ is
constrained to be
\begin{equation}
\left\langle \Delta\phi\right\rangle =\sum\limits_{n=1}^{N}\left\langle \Delta
x_{n}^{a}\right\rangle \frac{\partial\phi}{\partial x_{n}^{a}}=\kappa^{\prime
}~, \label{kappa prime}%
\end{equation}
where $\kappa^{\prime}$ is another small but for now unspecified
position-independent constant.

The introduction of the drift potential $\phi(x)$ will not be justified at
this point but, given its importance, a couple of brief comments may be of
value. First, we note that one can make progress by identifying the relevant
constraints even when their physical origin remains unexplained. This
situation is not unlike classical mechanics where identifying the forces can
be very useful even in those situations where their microscopic origin is not
yet fully understood. Second, it is in fact possible to provide a microscopic
interpretation of $\phi(x)$ in terms of the entropy of some other microscopic
variables that live at some deeper \textquotedblleft
sub-quantum\textquotedblright\ level \cite{Caticha 2010}. Whether postulating
such variables leads to further physical insights is an interesting topic for
future research.

\paragraph{The gauge constraints--}

The minimal assumptions described in the previous paragraphs already lead to
an interesting ED but we can construct richer forms of dynamics that describe
electromagnetic interactions by imposing additional constraints.

We assume that the motion of each particle $n$ is affected by an additional
field $\chi(x_{n})$ ($x_{n}\in\mathbf{X}$ is a point in physical 3D space)
with the topological properties of an angle: $\chi(x_{n})$ and $\chi
(x_{n})+2\pi$ describe the same angle. We further assume that these angles can
be redefined by different amounts $\gamma(x)$ at different places, that is,
the origin from which these angles are measured can be set independently at
each $x$. This is a local gauge symmetry and it immediately raises the
question of how can one compare angles at different locations. The answer is
well known: introduce a \emph{connection} field, a vector potential $A_{a}(x)$
that defines which angle at $x+\Delta x$ is the \textquotedblleft
same\textquotedblright\ as the angle at $x$. This is implemented by imposing
that as $\chi\rightarrow\chi+\gamma$ then the connection transforms as
$A_{a}\rightarrow A_{a}+\partial_{a}\gamma$ so that the corrected derivative
$\partial_{a}\chi-A_{a}$ remains invariant.\footnote{Note that since $\chi$ is
dimensionless the vector potential $A_{a}$ has units of inverse length. This
defines units of electric charge that differ from those conventionally adopted
in electromagnetism.} To derive an ED that incorporates gauge invariant
interactions for each particle $n$ we impose a constraint that involves these
corrected derivatives,
\begin{equation}
\langle\Delta x_{n}^{a}\rangle\left[  \partial_{na}\chi(x_{n})-A_{a}%
(x_{n})\right]  =\kappa_{n}^{\prime\prime}~,\quad(n=1\ldots N)~,
\label{constraint A}%
\end{equation}
where $\partial_{na}=\partial/\partial x_{n}^{a}$. The quantities $\kappa
_{n}^{\prime\prime}$ can be specified directly, but as is often the case in
entropic inference, it is much more convenient to specify the constraints
indirectly through the corresponding Lagrange multipliers $\beta_{n}$ which,
as we shall later see, will turn out to be the electric charges.

\paragraph*{The transition probability--}

The distribution $P(x^{\prime}|x)$ that maximizes the entropy $\mathcal{S}%
[P,Q]$ in (\ref{entropy a}) relative to (\ref{prior}) and subject to
(\ref{kappa prime}), (\ref{constraint A}), and normalization is
\begin{equation}
P(x^{\prime}|x)\propto\exp%
{\textstyle\sum\nolimits_{n}}
[-\frac{\alpha_{n}}{2}\delta_{ab}\Delta x_{n}^{a}\Delta x_{n}^{b}%
+\alpha^{\prime}\frac{\partial\phi}{\partial x_{n}^{a}}\Delta x_{n}^{a}%
+\beta_{n}\left(  \partial_{na}\chi_{n}-A_{a}(x_{n})\right)  \Delta x_{n}^{a}]
\label{trans prob a}%
\end{equation}
where $\alpha^{\prime}$ and $\beta_{n}$ are Lagrange multipliers. It is
convenient to rewrite $P(x^{\prime}|x)$ as
\begin{equation}
P(x^{\prime}|x)=\frac{1}{Z}\exp[-\frac{1}{2}%
{\textstyle\sum\nolimits_{n}}
\alpha_{n}\,\delta_{ab}\left(  \Delta x_{n}^{a}-\langle\Delta x_{n}^{a}%
\rangle\right)  \left(  \Delta x_{n}^{b}-\langle\Delta x_{n}^{b}%
\rangle\right)  ] \label{trans prob b}%
\end{equation}
where $Z$ is a normalization constant. Thus, a generic displacement $\Delta
x_{n}^{a}=x_{n}^{\prime a}-x_{n}^{a}$ can be expressed as an expected drift
plus a fluctuation,
\begin{equation}
\Delta x_{n}^{a}=\left\langle \Delta x_{n}^{a}\right\rangle +\Delta w_{n}%
^{a}\,\,,
\end{equation}
where
\begin{equation}
\langle\Delta x_{n}^{a}\rangle=\frac{1}{\alpha_{n}}\delta^{ab}\left[
\alpha^{\prime}\partial_{nb}\phi+\beta_{n}\partial_{nb}\chi(x_{n})-\beta
_{n}A_{b}(x_{n})\right]  ~~, \label{ED drift}%
\end{equation}%
\begin{equation}
\left\langle \Delta w_{n}^{a}\right\rangle =0\quad\text{and}\quad\langle\Delta
w_{n}^{a}\Delta w_{n}^{b}\rangle=\frac{1}{\alpha_{n}}\delta^{ab}~.
\label{ED fluctuations}%
\end{equation}
We see that for very short steps, as $\alpha_{n}\rightarrow\infty$, the
dynamics is dominated by fluctuations $\Delta w_{n}^{a}$ which are of order
$O(\alpha_{n}^{-1/2})$, while the drift $\langle\Delta x_{n}^{a}\rangle$ is
much smaller, only of order $O(\alpha_{n}^{-1})$. Thus, just as in Brownian
motion, the trajectory is continuous but not differentiable. In ED particles
have definite positions but their velocities are completely undefined. The
directionality of the motion and the correlations among the particles are
introduced by a systematic drift determined by the drift potential $\phi$ and
the gauge fields $\chi$ and $A_{a}$, while the particle fluctuations remain
isotropic and uncorrelated. We can also see that the effect of $\alpha
^{\prime}$ is to enhance or suppress the magnitude of the drift relative to
the fluctuations. The large $\alpha^{\prime}$ limit turns out to be the
Bohmian limit of ED (see \emph{e.g.}, \cite{Bartolomeo Caticha 2015}%
\cite{Bartolomeo Caticha 2016}). However, since the drift potential is at this
point unspecified, we can without loss of generality absorb $\alpha^{\prime}$
into $\phi$, $\alpha^{\prime}\phi\rightarrow\phi$, which amounts to setting
$\alpha^{\prime}=1$.

\section{Entropic time}

Having obtained a prediction, given by eq.(\ref{trans prob b}), for what
motion to expect in one infinitesimally short step we now consider motion over
finite distances. This requires the introduction of a parameter $t$, to be
called \emph{time}, as a book-keeping tool to keep track of the accumulation
of short steps. The construction of time involves three ingredients. First, we
must identify something that one might call an \textquotedblleft
instant\textquotedblright; second, it must be shown that these instants are
ordered; and finally, one must introduce a measure of separation between these
successive instants --- one must define \textquotedblleft
duration.\textquotedblright\ Since the foundation for any theory of time is
the theory of change --- that is, the underlying dynamics --- the notion of
time constructed below will reflect the inferential nature of entropic
dynamics. Such a construction we will call \emph{entropic time} \cite{Caticha
2010}.

\paragraph*{Time as an ordered sequence of instants--}

Entropic dynamics is given by a succession of short steps described by
$P(x^{\prime}|x)$, eq.(\ref{trans prob b}). Consider, for example, the $i$th
step which takes the system from $x=x_{i-1}$ to $x^{\prime}=x_{i}$.
Integrating the joint probability, $P(x_{i},x_{i-1})$, over $x_{i-1}$ gives
\begin{equation}
P(x_{i})=\int dx_{i-1}P(x_{i},x_{i-1})=\int dx_{i-1}P(x_{i}|x_{i-1}%
)P(x_{i-1})~. \label{CK a}%
\end{equation}
This equation follows directly from the laws of probability, it involves no
physical assumptions and, therefore, it is sort of empty. To make it useful
something else must be added. Suppose we interpret $P(x_{i-1})$ as the
probability of different values of $x_{i-1}$ \emph{at one \textquotedblleft
instant\textquotedblright\ labelled }$t$, then we can interpret $P(x_{i})$ as
the probability of values of $x_{i}$ \emph{at the next \textquotedblleft
instant\textquotedblright} which we will label $t^{\prime}$. Writing
$P(x_{i-1})=\rho_{t}(x)$ and $P(x_{i})=\rho_{t^{\prime}}(x^{\prime})$ we have
\begin{equation}
\rho_{t^{\prime}}(x^{\prime})=\int dx\,P(x^{\prime}|x)\rho_{t}(x)~.
\label{CK b}%
\end{equation}
Nothing in the laws of probability leading to eq.(\ref{CK a}) forces the
interpretation (\ref{CK b}) on us --- this is the additional ingredient that
allows us to construct time and dynamics in our model. Equation (\ref{CK b})
defines the notion of \textquotedblleft instant\textquotedblright:\ If the
distribution $\rho_{t}(x)$\ refers to one instant $t$, then the distribution
$\rho_{t^{\prime}}(x^{\prime})$\ generated by $P(x^{\prime}|x)$ defines what
we mean by the \textquotedblleft next\textquotedblright\ instant $t^{\prime}$.
The dynamics is defined by iterating this process. Entropic time is
constructed instant by instant: $\rho_{t^{\prime}}$ is constructed from
$\rho_{t}$, $\rho_{t^{\prime\prime}}$ is constructed from $\rho_{t^{\prime}}$,
and so on.

The construction is intimately related to information and inference. An
`instant' is an informational state that is complete in the sense that it is
specified by the information --- codified into the distributions $\rho_{t}(x)$
and $P(x^{\prime}|x)$ --- that is sufficient for predicting the next instant.
Thus, \emph{the present is defined such that,\ given the present, the future
is independent of the past}.

In the ED framework the notions of instant and of simultaneity are intimately
related to the distribution $\rho_{t}(x)$. It is instructive to discuss this
further. When we consider a single particle at a position $\vec{x}%
=(x^{1},x^{2},x^{3})$ in 3D space it is implicit in the notation that the
three coordinates $x^{1}$, $x^{2}$, and $x^{3}$ occur simultaneously --- no
surprises here. Things get a bit more interesting when we describe a system of
$N$ particles by a single point $x=(\vec{x}_{1},\vec{x}_{2},\ldots\vec{x}%
_{N})$ in $3N$-dimensional configuration space. Here it is also implicitly
assumed that all the $3N$ coordinate values refer to the same instant; we take
them to be simultaneous. There is an implicit assumption linking the very idea
of a point in configuration space with that of simultaneity. Whether we talk
about one particle or about $N$ particles, a distribution such as $\rho
_{t}(x)$ describes our uncertainty about the possible configurations $x$ of
the system at a given instant. Thus, in ED,\ a probability distribution
$\rho_{t}(x)$ provides a criterion of simultaneity.

In a relativistic theory there is a greater freedom in the choice of instants
and this translates into a greater flexibility with the notion of
simultaneity. Conversely, as we have shown elsewhere, the requirement that
these different notions of simultaneity be consistent with each other places
strict constraints on the allowed forms of relativistic ED \cite{Ipek Abedi
Caticha 2016}.

It is common to use equations such as (\ref{CK b}) to define a special kind of
dynamics, called Markovian, that unfolds in a time defined by some external
clocks. In such a Markovian dynamics the specification of the state at one
instant is sufficient to determine its evolution into the future. It is
important to emphasize that although the ED eq.(\ref{CK b}) is formally
identical to the Chapman-Kolmogorov equation we are not making a Markovian
assumption. We do not use (\ref{CK b}) to define a (Markovian) dynamics in a
pre-existing background time because in ED there are no external clocks. The
system provides its own clock and (\ref{CK b}) is used both to define the
dynamics and to construct time itself. In this respect, entropic time bears
some resemblance with the relational notion of time advocated by J. Barbour in
the context of classical physics (see \emph{e.g.} \cite{Barbour 1994}).

\paragraph*{The arrow of entropic time--}

The notion of time constructed according to eq.(\ref{CK b}) is intrinsically
directional. There is an absolute sense in which $\rho_{t}(x)$\ is prior and
$\rho_{t^{\prime}}(x^{\prime})$\ is posterior. If we wanted to construct a
time-reversed evolution we would write
\begin{equation}
\rho_{t}(x)=%
{\textstyle\int}
dx^{\prime}\,P(x|x^{\prime})\rho_{t^{\prime}}(x^{\prime})\,,
\end{equation}
where according to the rules of probability theory $P(x|x^{\prime})$ is
related to $P(x^{\prime}|x)$ in eq.(\ref{trans prob b}) by Bayes' theorem,
\begin{equation}
P(x|x^{\prime})=\frac{\rho_{t}(x)}{\rho_{t^{\prime}}(x^{\prime})}P(x^{\prime
}|x)~. \label{Bayes thm}%
\end{equation}
Note, however, that this is not a mere exchange of primed and unprimed
quantities. The distribution $P(x^{\prime}|x)$, eq.(\ref{trans prob b}), is a
Gaussian derived from the maximum entropy method. In contrast, the
time-reversed $P(x|x^{\prime})$ is given by Bayes' theorem,
eq.(\ref{Bayes thm}), and is not in general Gaussian. In (\ref{Bayes thm}) we
see that the asymmetry between the inferential past and the inferential future
arises from the asymmetry between priors and posteriors.

The puzzle of the arrow of time (see \emph{e.g.} \cite{Price 1996}\cite{Zeh
2002}) has been how to explain the asymmetric arrow from underlying symmetric
laws. The solution offered by ED is that there are no underlying laws whether
symmetric or not. The time asymmetry is the inevitable consequence of entropic
inference. From the point of view of ED the challenge is not to explain the
arrow of time but the reverse: how to explain the emergence of symmetric laws
within an entropic framework that is intrinsically asymmetric. As we shall see
below some laws of physics derived from ED, such as the Schr\"{o}dinger
equation, are indeed time-reversible even though entropic time itself is
strongly directional.

\paragraph*{Duration: a convenient time scale--}

To complete the construction of entropic time we need to specify the interval
$\Delta t$ between successive instants. The basic criterion is convenience:
\emph{duration is defined so that motion looks simple.} We saw in
eqs.(\ref{ED drift}) and (\ref{ED fluctuations}) that for short steps (large
$\alpha_{n}$) the motion is largely dominated by fluctuations. Therefore
specifying $\Delta t$ amounts to specifying the multipliers $\alpha_{n}$ in
terms of $\Delta t$.

The description of motion is simplest when it reflects the symmetry of
translations in space and time. In a flat spacetime this leads us to an
entropic time that resembles Newtonian time in that it flows \textquotedblleft
equably\ everywhere and everywhen.\textquotedblright\ Thus, we choose
$\alpha_{n}$ to be independent of $x$ and $t$, and we choose $\Delta t$ so
that $\alpha_{n}\propto1/\Delta t$. Furthermore, it is convenient to express
the proportionality constants in terms of some particle-specific constants
$m_{n}$ and an overall constant $\hbar$ that fixes the units of the $m_{n}$s
relative to the units of time. The result is
\begin{equation}
\alpha_{n}=\frac{m_{n}}{\hbar}\frac{1}{\Delta t}~. \label{alpha n}%
\end{equation}
As we shall see, the constants $m_{n}$ will eventually be identified with the
particle masses while the constant $\hbar$ will be identified as Planck's constant.

\section{The information metric of configuration space}

Before we proceed to study the dynamics defined eq.(\ref{trans prob b}) it is
useful to consider the geometry of the $N$-particle configuration space,
$\mathbf{X}_{N}$. Since the physical single particle space $\mathbf{X}$ is
described by the Euclidean metric $\delta_{ab}$ we can expect that the
$N$-particle configuration space, $\mathbf{X}_{N}=\mathbf{X}\times\ldots
\times\mathbf{X}$, will also be flat, but for non-identical particles a
question remains about the relative weights associated to each $\mathbf{X}$
factor. Information geometry provides the answer.

To each point $x\in\mathbf{X}_{N}$ there corresponds a probability
distribution $P(x^{\prime}|x)$. This means that to the configuration space
$\mathbf{X}_{N}$ we can associate a statistical manifold and its geometry is
uniquely determined (up to an overall scale factor) by the information metric,%
\begin{equation}
\gamma_{AB}=C\int dx^{\prime}\,P(x^{\prime}|x)\frac{\partial\log P(x^{\prime
}|x)}{\partial x^{A}}\frac{\partial\log P(x^{\prime}|x)}{\partial x^{B}}~.
\label{gamma C}%
\end{equation}
Here the upper case indices label both the particle and its coordinate,
$x^{A}=x_{n}^{a}$, and $C$ is an arbitrary positive constant (see \emph{e.g.},
\cite{Amari 1985}\cite{Caticha 2012}). Substituting eqs.(\ref{trans prob b})
and (\ref{alpha n}) into (\ref{gamma C}) in the limit of short steps
($\alpha_{n}\rightarrow\infty$) yields
\begin{equation}
\gamma_{AB}=\frac{Cm_{n}}{\hbar\Delta t}\delta_{nn^{\prime}}\,\delta
_{ab}=\frac{Cm_{n}}{\hbar\Delta t}\delta_{AB}~. \label{gamma AB}%
\end{equation}
The divergence as $\Delta t\rightarrow0$ arises because the information metric
measures statistical distinguishability. As $\Delta t\rightarrow0$ the
distributions $P(x^{\prime}|x)$ and $P(x^{\prime}|x+\Delta x)$ become more
sharply peaked and increasingly easier to distinguish. Therefore, $\gamma
_{AB}\rightarrow\infty$. To define a geometry that remains useful even for
arbitrarily small $\Delta t$ we choose $C\propto\Delta t$. In fact, since
$\gamma_{AB}$ will always appear in the combination $\gamma_{AB}\Delta t/C$,
it is best to absorb these constants into a \textquotedblleft
mass\textquotedblright\ tensor,
\begin{equation}
m_{AB}=\frac{\hbar\Delta t}{C}\gamma_{AB}=m_{n}\delta_{AB}~. \label{mass a}%
\end{equation}
The inverse mass tensor will also turn out to be useful,
\begin{equation}
m^{AB}=\frac{C}{\hbar\Delta t}\gamma^{AB}=\frac{1}{m_{n}}\delta^{AB}~.
\label{mass b}%
\end{equation}

Thus, up to overall constants the mass tensor is the metric of configuration
space. Ever since the work of Heinrich Hertz in 1894 \cite{Lanczos 1970} it
has been standard practice to describe the motion of systems with many
particles as the motion of a single point in an abstract space --- the
configuration space. Choosing the geometry of this configuration space based
on an examination of the kinetic energy of the system has so far been regarded
as a matter of convenience. In ED there is no choice: up to a global scale
factor the metric follows uniquely from information geometry.

To summarize our results so far: with the constants $\alpha_{n}$ chosen
according to (\ref{alpha n}), the dynamics given by $P(x^{\prime}|x)$
in\ (\ref{trans prob b}) is a standard Wiener process. A generic displacement
is written as a drift plus a fluctuation,
\begin{equation}
\Delta x^{A}=b^{A}\Delta t+\Delta w^{A}~. \label{Delta x}%
\end{equation}
From eq.(\ref{ED drift}) the drift velocity is
\begin{equation}
b^{A}=\frac{\langle\Delta x^{A}\rangle}{\Delta t}=m^{AB}\left[  \partial
_{B}\hbar\left(  \phi+\bar{\chi}\right)  -\bar{A}_{B}\right]  ~,
\label{drift velocity}%
\end{equation}
where we have introduced the configuration space quantities,
\begin{equation}
\bar{\chi}(x)=%
{\textstyle\sum\nolimits_{n}}
\beta_{n}\chi(x_{n})\quad\text{and}\quad\bar{A}_{A}(x)=\hbar\beta_{n}%
A_{a}(x_{n})~, \label{phibar Abar}%
\end{equation}
and the fluctuation $\Delta w^{A}$ is given by
\begin{equation}
\langle\Delta w^{A}\rangle=0\quad\text{and}\quad\langle\Delta w^{A}\Delta
w^{B}\rangle=\hbar m^{AB}\Delta t~. \label{fluc}%
\end{equation}

I finish this section with two remarks. The first is on \emph{the nature of
clocks}: In Newtonian mechanics time is defined to simplify the motion of free
particles. The prototype of a clock is a free particle which moves equal
distances in equal times. In ED time is also defined to simplify the dynamics
of free particles (for sufficiently short times all particles are free) and
the prototype of a clock is a free particle too: as we see in (\ref{fluc})
\emph{a free particle's variance increases by equal amounts in equal times}.

The second remark is on \emph{the nature of mass}. In standard quantum
mechanics, \textquotedblleft what is mass?\textquotedblright\ and
\textquotedblleft why quantum fluctuations?\textquotedblright\ are two
independent mysteries. In ED the mystery is somewhat alleviated: mass and
fluctuations are two sides of the same coin. \emph{Mass is an inverse measure
of fluctuations}.

\section{Diffusive dynamics}

The dynamics of $\rho(x)$, given by the integral equation (\ref{CK b}), is
more conveniently re-written in a differential form known as the Fokker-Planck
(FP) equation,
\begin{equation}
\partial_{t}\rho=-\partial_{A}\left(  b^{A}\rho\right)  +\frac{1}{2}\hbar
m^{AB}\partial_{A}\partial_{B}\rho~. \label{FP a}%
\end{equation}
(For the algebraic details see \emph{e.g.}, \cite{Caticha 2012}.) The FP
equation can also be written as a continuity equation,
\begin{equation}
\partial_{t}\rho=-\partial_{A}\left(  \rho v^{A}\right)  ~. \label{FP b}%
\end{equation}
The product $\rho v^{A}$ in (\ref{FP b}) represents the probability current,
and $v^{A}$ is interpreted as the velocity of the probability flow --- it is
called the \emph{current velocity}. From (\ref{FP a}) the current velocity in
(\ref{FP b}) is the sum of two separate contributions,
\begin{equation}
v^{A}=b^{A}+u^{A}\,. \label{curr a}%
\end{equation}
where $u^{A}$ is the \emph{osmotic velocity},
\begin{equation}
u^{A}=-\hbar m^{AB}\partial_{B}\log\rho^{1/2}~.
\end{equation}
It represents diffusion, the tendency for probability to flow down the density
gradient. Since both $b^{A}$ and $u^{A}$ involve gradients the current
velocity can be written in gauge invariant form,
\begin{equation}
v^{A}=m^{AB}(\partial_{B}\Phi-\bar{A}_{B})\quad\text{where}\quad\Phi
=\hbar(\phi+\bar{\chi}-\log\rho^{1/2})~. \label{curr b}%
\end{equation}
The field $\Phi$, which we will call the \textquotedblleft
phase\textquotedblright, plays a central role in what follows. The FP equation
(\ref{FP b}) can be conveniently rewritten in yet another equivalent but very
suggestive form involving functional derivatives. For some suitably chosen
functional $\tilde{H}[\rho,\Phi]$ we have
\begin{equation}
\partial_{t}\rho(x)=-\partial_{A}\left[  \rho m^{AB}(\partial_{B}\Phi-\bar
{A}_{B})\right]  =\frac{\delta\tilde{H}}{\delta\Phi(x)}~. \label{Hamilton a}%
\end{equation}
It is easy to check that the appropriate functional $\tilde{H}$ is%
\begin{equation}
\tilde{H}[\rho,\Phi]=\int dx\,\frac{1}{2}\rho m^{AB}\left(  \partial_{A}%
\Phi-\bar{A}_{A}\right)  \left(  \partial_{B}\Phi-\bar{A}_{B}\right)
+F[\rho]~, \label{Hamiltonian a}%
\end{equation}
where the unspecified functional $F[\rho]$ is an integration constant.

Equation (\ref{Hamilton a}) describes a standard diffusion involving a single
dynamical field $\rho(x)$ that evolves in response to the non-dynamical
fields\ $\phi$, $\bar{\chi}$, and $\bar{A}$. In contrast, a \emph{quantum}
dynamics consists in the coupled evolution of two dynamical fields: the
density $\rho(x)$ and the phase of the wave function. This second field can be
naturally introduced into ED by allowing the field $\Phi$ in (\ref{curr b}) to
become dynamical:\ the phase $\Phi$ guides the evolution of $\rho$, and in
return, the evolving $\rho$ reacts back and induces a change in $\Phi$. This
amounts to an ED in which the constraint (\ref{kappa prime}) is continuously
updated at each instant in time. To find the appropriate updating criterion,
which gives the dynamics in e-phase space we appeal once again to information geometry.

\section{The geometry of e-phase space}

It may be convenient to recall one particular derivation of the information
metric. In the discrete case the statistical manifold is the simplex
$\mathcal{S}_{\nu-1}=\{\rho=(\rho^{1}\ldots\rho^{\nu}):%
{\textstyle\sum_{i}^{\nu}}
\rho^{i}=1\}$. As is common in the context of probability and inference we
appeal to symmetry. Changing to new coordinates $\xi^{i}=(\rho^{i})^{1/2}$ the
equation for the simplex $\mathcal{S}_{\nu-1}$ --- the normalization condition
--- reads $%
{\textstyle\sum_{i}^{\nu}}
(\xi^{i})^{2}=1$ which suggests the equation of a sphere. We take this hint
seriously and\emph{ declare} that the symplex is an $(\nu-1)$-sphere embedded
in an $\nu$-dimensional spherically symmetric space.\footnote{One might be
concerned that this particular choice of symmetry is an ad hoc assumption but
the result is very robust. Exactly the same metric is obtained by several
different criteria \cite{Campbell 1986}\cite{Caticha 2012}.} The metric of a
generic spherically symmetric space takes the form
\begin{equation}
d\ell^{2}=\left(  a(|\rho|)-b(|\rho|)\right)  \left(
{\textstyle\sum\nolimits_{i}^{\nu}}
\xi^{i}d\xi^{i}\right)  ^{2}+|\rho|{}b(|\rho|{})%
{\textstyle\sum\nolimits_{i}^{\nu}}
(d\xi^{i})^{2}, \label{info metric a}%
\end{equation}
where $a(|\rho|)$ and $b(|\rho|)$ are two arbitrary smooth and positive
functions of $|\rho|{}=%
{\textstyle\sum\nolimits_{i}^{\nu}}
\rho^{i}$. Changing back to the original $\rho^{i}$ coordinates and setting
$|\rho|{}=1$ and $%
{\textstyle\sum\nolimits_{i}^{\nu}}
d\rho^{i}=0$~gives the information metric up to an overall scale,
\begin{equation}
d\ell^{2}=b(1)%
{\textstyle\sum\nolimits_{i}^{\nu}}
\frac{1}{\rho^{i}}(d\rho^{i})^{2}~. \label{info metric b}%
\end{equation}

To extend the information metric from the symplex $\mathcal{S}_{\nu-1}$ to the
$2\nu$-dimensional e-phase space $(\rho^{i},\Phi^{i})$ we impose two
conditions: (A) that the extended metric be compatible with the information
metric on the symplex $\mathcal{S}_{\nu-1}$, and (B) that $\Phi$ inherits from
the gauge potential $\chi$ the topological structure of an angle. The simplest
way to implement (A) is to follow as closely as possible the derivation that
led to (\ref{info metric b}). Condition (B) suggests introducing coordinates,
\begin{equation}
\xi^{i}=(\rho^{i})^{1/2}\cos\Phi^{i}/\hbar\quad\text{and }\quad\eta^{i}%
=(\rho^{i})^{1/2}\sin\Phi^{i}/\hbar~.
\end{equation}
Then the normalization condition reads
\begin{equation}
|\rho|{}=%
{\textstyle\sum\nolimits_{i}^{\nu}}
\rho^{i}=%
{\textstyle\sum\nolimits_{i}^{\nu}}
\left[  (\xi^{i})^{2}+(\eta^{i})^{2}\right]  =1
\end{equation}
which suggests the equation of a sphere in $2\nu$ dimensions and, once again,
we take the spherical symmetry seriously. The most general metric in the space
$\{\rho,\Phi\}$ that is invariant under rotations is
\begin{equation}
d\ell^{2}=\left(  a(|\rho|{})-b(|\rho|{})\right)  \left[
{\textstyle\sum\nolimits_{i}^{\nu}}
\left(  \xi^{i}d\xi^{i}+\eta^{i}d\eta^{i}\right)  \right]  ^{2}+|\rho
|{}b(|\rho|{})%
{\textstyle\sum\nolimits_{i}^{\nu}}
\left[  (d\xi^{i})^{2}+(d\eta^{i})^{2}\right]  \,.
\end{equation}
Transforming back to the coordinates $(\rho^{i},\Phi^{i})$, setting $|\rho
|{}=1$ and $%
{\textstyle\sum\nolimits_{i}^{\nu}}
d\rho^{i}=0$, and dropping an unimportant proportionality constant, leads to
\begin{equation}
d\ell^{2}=\sum\nolimits_{i}^{\nu}\left[  \frac{\hbar}{2\rho^{i}}(d\rho
^{i})^{2}+\frac{2}{\hbar}\rho^{i}(d\Phi^{i})^{2}\right]  \,.
\end{equation}
Finally, generalizing to the continuous e-phase space gives the desired
result,
\begin{equation}
\delta\ell^{2}=\int dx\,\left[  \frac{\hbar}{2\rho}\delta\rho^{2}+\frac
{2}{\hbar}\rho\delta\Phi^{2}\right]  ~. \label{eps metric a}%
\end{equation}

\paragraph*{Index notation--}

To deal with tensors in e-phase space we introduce a convenient index
notation. A point $X$ will be labelled by its coordinates $X^{\alpha x}$,
where $\alpha$ takes one of two values, $\rho$ or $\Phi$. Thus,
\begin{equation}
X^{\alpha x}=\left(  X^{\rho x},X^{\Phi x}\right)  =\left(  \rho
(x),\Phi(x)\right)  ~.
\end{equation}
Repeated indices will as usual imply sums and integrals. For example,
eq.(\ref{eps metric a}) is written as
\begin{equation}
\delta\ell^{2}=\int dxdx^{\prime}G_{\alpha\alpha^{\prime}}(x,x^{\prime})\delta
X^{\alpha x}\delta X^{\alpha^{\prime}x^{\prime}}=G_{\alpha x,\beta x^{\prime}%
}\delta X^{\alpha x}\delta X^{\beta x^{\prime}}%
\end{equation}
where the metric tensor $G$ has components $G_{\alpha x,\beta x^{\prime}}$,
\begin{equation}
G_{\alpha x,\beta x^{\prime}}=G_{\alpha\beta}(x,x^{\prime})=%
\begin{bmatrix}
\hbar/2\rho(x) & 0\\
0 & 2\rho(x)/\hbar
\end{bmatrix}
\delta(x,x^{\prime})~. \label{eps metric b}%
\end{equation}
The scalar product of two vectors $V^{\alpha x}$ and $U^{\beta x^{\prime}}$
is
\begin{equation}
G[V,U]=G_{\alpha x,\beta x^{\prime}}V^{\alpha x}U^{\beta x^{\prime}}~.
\end{equation}

\paragraph*{Complex and Symplectic structures--}

A linear transformation $V\rightarrow JV=V_{J}$ or, more explicitly,
\begin{equation}
V_{J}^{\alpha x}=J^{\alpha x}{}_{\beta x^{\prime}}V^{\beta x^{\prime}}=\int
dx^{\prime}J^{\alpha}{}_{\beta}(x,x^{\prime})V^{\beta x^{\prime}}~,
\end{equation}
is said to be a symmetry of the metric $G$, and we say that $G$ is
$J$-invariant, when
\begin{equation}
G[V_{J},U_{J}]=G[V,U]~
\end{equation}
for all vectors $V$ and $U$. Equivalently, $G$ is $J$-invariant when
\begin{equation}
G_{\alpha x,\beta x^{\prime}}J^{\alpha x}{}_{\gamma x^{\prime\prime}}J^{\beta
x^{\prime}}{}_{\delta x^{\prime\prime\prime}}=G_{\gamma x^{\prime},\delta
x^{\prime\prime\prime}}~. \label{J invariance}%
\end{equation}

It turns out that, beyond spherical symmetry, the metric (\ref{eps metric b})
has another crucial symmetry. It is invariant under the transformation
\begin{equation}
J^{\alpha x}{}_{\beta x^{\prime}}=J^{\alpha}{}_{\beta}(x,x^{\prime})=%
\begin{bmatrix}
0 & 2\rho(x)/\hbar\\
-\hbar/2\rho(x) & 0
\end{bmatrix}
\delta(x,x^{\prime})~.~~ \label{J}%
\end{equation}
The proof is straightforward: just substitute (\ref{J}) to verify it satisfies
(\ref{J invariance}).

The importance of $J$ invariance is twofold. First, applying $J$ twice gives
$J^{2}=-\mathbf{1}$, or%
\begin{equation}
J^{\alpha x}{}_{\beta x^{\prime}}J^{\beta x^{\prime}}{}_{\gamma x^{\prime
\prime}}=-\delta^{\alpha x}{}_{\gamma x^{\prime\prime}}=-\delta^{\alpha}%
{}_{\gamma}\delta(x,x^{\prime\prime})~. \label{J complex}%
\end{equation}
Therefore the metric (\ref{eps metric a}) automatically endows e-phase space
with a complex structure $J^{\alpha x}{}_{\beta x^{\prime}}$. This is the
reason why complex numbers play such a central role of quantum mechanics.
Second, we can use the metric tensor to lower the first index in $J$,
\begin{equation}
G_{\alpha x,\beta x^{\prime}}J^{\beta x^{\prime}}{}_{\gamma x^{\prime\prime}%
}=J_{\beta x^{\prime},\gamma x^{\prime\prime}}~.
\end{equation}
The result is the antisymmetric tensor,
\begin{equation}
\Omega_{\alpha x,\beta x^{\prime}}=J_{\alpha x,\beta x^{\prime}}=%
\begin{bmatrix}
0 & 1\\
-1 & 0
\end{bmatrix}
\delta(x,x^{\prime})~. \label{symplectic}%
\end{equation}
Thus e-phase space is also automatically endowed with a symplectic 2-form
$\Omega$.\footnote{It is easy to check that $\Omega$ is also $J$-invariant:
$\Omega\lbrack V_{J},U_{J}]=\Omega\lbrack V,U]$ for all vectors $V$ and $U$.}
This is what justifies calling it a phase space in the canonical Hamiltonian
sense of the term\emph{. }

It is worth emphasizing that the complex $J$ and symplectic $\Omega$
structures have been derived and not imposed. The extended metric was derived
from premises that are natural in a probabilistic setting (information
geometry, spherical symmetry, the topological nature of the field $\Phi$)
without any reference to principles of mechanics.

Of course, the fact that our premises imply the symplectic and complex
structures means that the latter where somehow already implicit in the former.
One might therefore consider an alternative approach that inverts the logic:
start by assuming the existence of complex and symplectic structures and from
these attempt to derive the metric (\ref{eps metric a}) --- but this approach
has several drawbacks. One is that it would prevent us from gaining any deeper
insight into why do symplectic structures arise in physics in the first place.
Another is that the metric derived in this way is not unique; the alternative
approach leads to a continuum of metrics of which (\ref{eps metric a}) is a
member \cite{Reginatto Hall 2012}. Our premises are more restrictive than
merely imposing complex and symplectic structures.

\paragraph*{Poisson brackets--}

The availability of the symplectic 2-form $\Omega_{\alpha x,\beta x^{\prime}}$
leads to the canonical Hamilton-Jacobi formalism. For example, by raising
indices one constructs a tensor $\Omega^{\alpha x,\beta x^{\prime}}$ which, as
can be easily checked, has the same components as (\ref{symplectic}). Indeed,
\begin{equation}
\Omega^{\alpha x,\beta x^{\prime}}\Omega_{\beta x^{\prime},\gamma
x^{\prime\prime}}=\Omega^{\alpha x}{}_{\beta x^{\prime}}\Omega^{\beta
x^{\prime}}{}_{\gamma x^{\prime\prime}}=-\delta^{\alpha x}{}_{\gamma
x^{\prime\prime}}~.
\end{equation}
To see its utility consider the gradient 1-form $\tilde{\nabla}F$ of a generic
functional $F[\rho,\Phi]$. Its components are
\begin{equation}
\tilde{\nabla}_{\alpha x}F=\left(  \frac{\delta F}{\delta\rho(x)},\frac{\delta
F}{\delta\Phi(x)}\right)  ~.
\end{equation}
The action of $\Omega^{\alpha x,\beta x^{\prime}}$ on any two 1-forms
$\tilde{\nabla}F_{1}$ and $\tilde{\nabla}F_{2}$ is the Poisson bracket,
\begin{equation}
\Omega^{\alpha x,\beta x^{\prime}}\tilde{\nabla}_{\alpha x}F_{1}\tilde{\nabla
}_{\beta x^{\prime}}F_{2}=\int dx\left(  \frac{\delta F_{1}}{\delta\rho
(x)}\frac{\delta F_{2}}{\delta\Phi(x)}-\frac{\delta F_{2}}{\delta\rho(x)}%
\frac{\delta F_{1}}{\delta\Phi(x)}\right)  =\{F_{1},F_{2}\}~. \label{PB a}%
\end{equation}
The special case $F_{1}=\rho$ and $F_{2}=\Phi$,
\begin{equation}
\{\rho(x),\Phi(x^{\prime})\}=\delta(x,x^{\prime})~,\quad\text{and\quad}%
\{\rho(x),\rho(x^{\prime})\}=\{\Phi(x),\Phi(x^{\prime})\}=0~, \label{PB b}%
\end{equation}
shows that $\rho(x)$ and $\Phi(x)$ are canonically conjugate variables.

\paragraph*{Complex coordinates--}

The availability of a complex structure suggests introducing complex
coordinates
\begin{equation}
\Psi=\rho^{1/2}\exp(i\Phi/\hbar)\quad\text{and}\quad\Psi^{\ast}=\rho^{1/2}%
\exp(-i\Phi/\hbar)~. \label{complex coords}%
\end{equation}
Indeed, the metric (\ref{eps metric a}) takes a very simple form,
\begin{equation}
\delta\ell^{2}=2\hbar\int dx\,\delta\Psi^{\ast}\delta\Psi~.
\label{eps metric c}%
\end{equation}
The corresponding Poisson brackets,
\begin{equation}
\{\Psi(x),\Psi^{\ast}(x^{\prime})\}=\frac{1}{i\hbar}\delta(x,x^{\prime
})~,\quad\text{and\quad}\{\Psi(x),\Psi(x^{\prime})\}=\{\Psi^{\ast}%
(x),\Psi^{\ast}(x^{\prime})\}=0~,
\end{equation}
show that $i\hbar\Psi^{\ast}(x)$ is the momentum that is canonically conjugate
to the complex coordinate $\Psi(x)$.

\section{The e-Hamiltonian}

According to entropic dynamics the evolution of $\rho$ is given by the FP
eq.(\ref{Hamilton a}). Now we seek a criterion to update the phase field
$\Phi$ --- which amounts to updating the constraints (\ref{kappa prime}) and
(\ref{constraint A}). The joint dynamics of $\rho$ and $\Phi$ must reproduce
the FP eq.(\ref{Hamilton a}); it is also natural to require that it preserve
the metric, complex, and symplectic structures of e-phase space. Such a
dynamics is a Hamiltonian flow in e-phase space.

The construction of the ensemble Hamiltonian --- or \emph{e-Hamiltonian} ---
involves three ingredients. First, if temporal distance has to do with whether
we can detect a change from one state to another. The question is: Does a
small change make any difference? Or, can we distinguish the new pair
$[\rho+\delta\rho,\Phi+\delta\Phi]$ from the old $[\rho,\Phi]$? This is
precisely the kind of issue that the metric (\ref{eps metric a}) has been
designed to address. Then it is only natural that the generator of change be
in some way related to the e-phase metric. Second, in ED entropic time is
constructed so that time (duration) is defined by a clock that is provided by
the system itself. More precisely, time is measured by the fluctuations of
free particles. It is only natural to require that the generator of time
translations be defined in terms of the same clock. And third, the
e-Hamiltonian has to agree with (\ref{Hamiltonian a}) in order to reproduce
the ED evolution of $\rho$, eq.(\ref{Hamilton a}).

To implement the first condition we note that in ED position plays the
privileged role of being the only ontic variable. We will therefore consider a
small change of state $\{\delta\rho,\delta\Phi\}$ when the position is
slightly shifted by $\delta x^{A}(x)$,
\begin{equation}
\delta\rho=\partial_{A}\rho\,\delta x^{A}\,, \label{displ rho}%
\end{equation}
The change in $\Phi$ is given by a similar expression except that in order to
compare neighboring phases we have to keep track of the gauge connection
$A_{A}$. The gauge invariant change in $\Phi$ is
\begin{equation}
\delta\Phi=\left(  \partial_{A}\Phi+\bar{A}_{A}\right)  \,\delta x^{A}~.
\label{displ Phi}%
\end{equation}
Substituting into (\ref{eps metric a}) gives,
\begin{equation}
\delta\ell^{2}={}\frac{4}{\hbar}\int dx\,h_{AB}\delta x^{A}\delta x^{B}~,
\label{hAB a}%
\end{equation}
where
\begin{equation}
h_{AB}=\frac{1}{2}\rho\left(  \partial_{A}\Phi-\bar{A}_{A}\right)  \left(
\partial_{B}\Phi-\bar{A}_{B}\right)  +\frac{\hbar^{2}}{8\rho}\partial_{A}%
\rho\partial_{B}\rho~. \label{hAB b}%
\end{equation}
Up to the irrelevant factor of $4/\hbar$, the tensor $h_{AB}$ is the induced
metric in configuration space --- it measures distinguishability under small displacements.

The second requirement --- that duration be measured by fluctuations --- is
implemented by demanding that the shift $\delta x^{A}$ be associated to
fluctuations in a short time interval $\Delta t$.\footnote{I am grateful to S.
Ipek and K. Vanslette for a very useful suggestions on this issue.} Recalling
(\ref{trans prob b}) and (\ref{fluc}), we see that as $\Delta t\rightarrow0$,
the product $\delta x^{A}\delta x^{B}$ converges \emph{in probability} to
$\langle\delta x^{A}\delta x^{B}\rangle$,
\begin{equation}
\delta x^{A}\delta x^{B}\sim\left\langle \delta x^{A}\delta x^{B}\right\rangle
+o(\Delta t)=\hbar m^{AB}\Delta t~\delta^{ab}+o(\Delta t)~. \label{dx2}%
\end{equation}
(See \cite{Nelson 1985}, section \S 5.) Therefore,
\begin{equation}
\delta\ell^{2}\sim{}4\Delta t\,\int dx\,m^{AB}h_{AB}+o(\Delta t)~. \label{dl2}%
\end{equation}
To satisfy the third requirement, we compare (\ref{Hamiltonian a}) with
(\ref{hAB b}) and (\ref{dl2}). This suggests an e-Hamiltonian,
\begin{equation}
\tilde{H}_{0}=\int dx\,m^{AB}h_{AB}~,
\end{equation}
that incorporates all the desired symmetries and even includes the quantum
potential term with the right coefficient. However, since the argument relies
on the short time evolution --- the dynamics of free particles --- the
construction cannot be expected to reproduce an interaction terms such as a
standard scalar potential $V(x)$. Therefore to account for the possibility of
additional interactions we introduce an additional potential term
$V(x)$,\footnote{A scalar potential such as $\rho(x)V(x)$ would arise
naturally in a model in which the electromagnetic field is represented by
dynamical degrees of freedom.}
\begin{equation}
\tilde{H}[\rho,\Phi]=\int dx\,\rho\left[  \frac{1}{2}m^{AB}\left(
\partial_{A}\Phi-\bar{A}_{A}\right)  \left(  \partial_{B}\Phi-\bar{A}%
_{B}\right)  +\frac{\hbar^{2}}{8\rho^{2}}m^{AB}\partial_{A}\rho\partial
_{B}\rho+V(x)\right]  . \label{Hamiltonian b}%
\end{equation}

The corresponding Hamilton equations reproduce the FP eq.(\ref{Hamilton a}),
\begin{equation}
\partial_{t}\rho=\frac{\delta\tilde{H}}{\delta\Phi}=-\partial_{A}\left[  \rho
m^{AB}\left(  \partial_{B}\Phi-\bar{A}_{B}\right)  \right]  ~, \label{FP c}%
\end{equation}
and the analogue of the Hamilton-Jacobi equation,
\begin{equation}
\partial_{t}\Phi=-\frac{\delta\tilde{H}}{\delta\rho}=-\frac{1}{2}m^{AB}\left(
\partial_{A}\Phi-\bar{A}_{A}\right)  \left(  \partial_{B}\Phi-\bar{A}%
_{B}\right)  +\frac{\hbar^{2}}{2}m^{AB}\frac{\partial_{A}\partial_{B}%
\rho^{1/2}}{\rho^{1/2}}-V\,. \label{HJ}%
\end{equation}
We can now combine $\rho$ and $\Phi$ into the wave function, $\Psi=\rho
^{1/2}\exp(i\Phi/\hbar)$. Expressing (\ref{FP c}) and (\ref{HJ}) in terms of
$\Psi$ leads to the Schr\"{o}dinger equation.

However, there is an equivalent but simpler derivation: use complex
coordinates (\ref{complex coords}) from the start. Use (\ref{displ rho}) and
(\ref{displ Phi}) to find
\begin{equation}
\delta\Psi=D_{A}\Psi\,\delta x^{A}~,
\end{equation}
where
\begin{equation}
D_{A}=\partial_{A}-\frac{i}{\hbar}\bar{A}_{A}=D_{na}=\frac{\partial}{\partial
x_{n}^{a}}-i\beta_{n}A_{a}(x_{n})~.
\end{equation}
Substitute into (\ref{eps metric c}), and use (\ref{dx2}) to get,
\begin{equation}
\delta\ell^{2}\sim4\Delta t\int dx\,\frac{\hbar^{2}}{2}m^{AB}(D_{A}\Psi
)^{\ast}D_{B}\Psi+o(\Delta t)~,
\end{equation}
which leads to the same e-Hamiltonian as eq.(\ref{Hamiltonian b}),
\begin{equation}
\tilde{H}[\Psi,\Psi^{\ast}]=\int dx\,\Psi^{\ast}\left[  -\frac{\hbar^{2}}%
{2}m^{AB}D_{A}D_{B}+V(x)\right]  \Psi\label{Hamiltonian c}%
\end{equation}
The Hamilton equation,
\begin{equation}
\partial_{t}\Psi(x)=\frac{\delta\tilde{H}}{\delta(i\hbar\Psi^{\ast}(x))}~,
\end{equation}
is the Schr\"{o}dinger equation,
\begin{equation}
i\hbar\partial_{t}\Psi=-\frac{\hbar^{2}}{2}m^{AB}D_{A}D_{B}\Psi+V\Psi
~.\label{sch a}%
\end{equation}
In more standard notation it reads
\begin{equation}
i\hbar\partial_{t}\Psi=-%
{\displaystyle\sum\nolimits_{n}}
\frac{\hbar^{2}}{2m_{n}}\delta^{ab}D_{na}D_{nb}\Psi+V\Psi~.\label{sch b}%
\end{equation}
At this point we can finally provide the physical interpretation of the
various constants introduced along the way: $\hbar$ is Planck's constant,
$m_{n}$ are the particles' masses, and the $\beta_{n}$ are related to the
particles' electric charges $q_{n}$ by%
\begin{equation}
\beta_{n}=\frac{q_{n}}{\hbar c}~.\label{charge a}%
\end{equation}

\section{On linearity, single-valuedness, and charge \newline quantization}

Earlier we remarked on the tension between the probabilistic and linear
structures in QM. As described by eqs.(\ref{FP c}) and (\ref{HJ}), ED is a
fully probabilistic theory and its representation in terms of the
Schr\"{o}dinger equation (\ref{sch b}) is clearly linear. However, this does
not yet guarantee the full compatibility of probability with linearity. It is
clear that if $\Psi_{1}$ and $\Psi_{2}$ are two solutions of eq.(\ref{sch b})
then the linear combination
\begin{equation}
\Psi_{3}=\alpha_{1}\Psi_{1}+\alpha_{2}\Psi_{2}%
\end{equation}
is also a solution. The problem is that even when $|\Psi_{1}|^{2}=\rho_{1}$
and $|\Psi_{2}|^{2}=\rho_{2}$ are probabilities, it is not generally true that
$|\Psi_{3}|^{2}$ will be a probability too.

To see the difficulty note that for generic choices of $\beta_{n}$ the wave
functions $\Psi$ are in general multi-valued. As one moves in a closed loop
$\Gamma$ in configuration space the wave function changes by a phase factor,
\begin{equation}
\Psi\rightarrow\Psi^{\prime}=e^{i\delta}\Psi~. \label{psi delta}%
\end{equation}
The superposition of $\Psi_{1}$ and $\Psi_{2}$ is multi-valued too,
\begin{equation}
\Psi_{3}\rightarrow\Psi_{3}^{\prime}=\alpha_{1}e^{i\delta_{1}}\Psi_{1}%
+\alpha_{2}e^{i\delta_{2}}\Psi_{2}~.
\end{equation}
The problem is that even if the magnitudes $|\Psi_{1}|^{2}=\rho_{1}$ and
$|\Psi_{2}|^{2}=\rho_{2}$ are single-valued (because they are probability
densities), $|\Psi_{3}|^{2}$\ will in general turn out to be multi-valued.
Indeed,
\begin{equation}
|\Psi_{3}|^{2}=|\alpha_{1}|^{2}\rho_{1}+|\alpha_{2}|^{2}\rho_{2}%
+2\operatorname{Re}[\alpha_{1}\alpha_{2}^{\ast}\Psi_{1}\Psi_{2}^{\ast}]~,
\end{equation}
changes into
\begin{equation}
|\Psi_{3}^{\prime}|^{2}=|\alpha_{1}|^{2}\rho_{1}+|\alpha_{2}|^{2}\rho
_{2}+2\operatorname{Re}[\alpha_{1}\alpha_{2}^{\ast}e^{i(\delta_{1}-\delta
_{2})}\Psi_{1}\Psi_{2}^{\ast}]~,
\end{equation}
so that in general
\begin{equation}
|\Psi_{3}^{\prime}|^{2}\neq|\Psi_{3}|^{2}~,
\end{equation}
which precludes the interpretation of $|\Psi_{3}|^{2}$ as a probability. The
problem does not arise if the phase $\delta$ in (\ref{psi delta}) has the same
value for all wave functions. Furthermore, since for some wave functions we
have $\delta=0$, it must be that $\delta=0$ for all wave functions. Therefore,
\emph{the condition of compatibility of the probabilistic and linear
structures is that the wave functions be single-valued.}

The single-valuedness condition is%
\begin{equation}
\Delta\frac{\Phi}{\hbar}=%
{\displaystyle\oint_{\Gamma}}
d\ell^{A}\partial_{A}\frac{\Phi}{\hbar}=2\pi\nu(\Gamma)~, \label{circ}%
\end{equation}
where $\nu(\Gamma)$ is an integer. To determine the choice of constraints ---
that is, to determine the values of the $\beta_{n}$ --- that lead to
single-valuedness consider a closed loop $\Gamma_{n}$ in which all particles
but the $n$th are kept fixed. Since $\chi$ is an angle,
\begin{equation}
\Delta\chi=%
{\displaystyle\oint_{\Gamma_{n}}}
d\ell_{n}^{a}\partial_{na}\chi=2\pi\nu(\Gamma_{n})~,
\end{equation}
where $\nu(\Gamma_{n})$ is an integer that depends on the loop $\Gamma_{n}$.
Furthermore, since $\phi(x)$ and $\log\rho(x)$ are single-valued, from
(\ref{curr b}), we have
\begin{equation}
\Delta\frac{\Phi}{\hbar}=%
{\displaystyle\oint_{\Gamma_{n}}}
d\ell_{n}^{a}\partial_{na}\frac{\Phi}{\hbar}=\beta_{n}%
{\displaystyle\oint_{\Gamma_{n}}}
d\ell_{n}^{a}\partial_{na}\chi=2\pi\nu(\Gamma_{n})\beta_{n}~.
\end{equation}
Comparing with eq.(\ref{circ}) for any arbitrary loops shows that $\beta_{n}$
must be an integer. Therefore, the choice of $\kappa_{n}$ in the gauge
constraints (\ref{constraint A}) that leads to compatibility of probabilistic
and linear structures is that the corresponding Lagrange multipliers
$\beta_{n}$ be integers.

\paragraph*{Quantized electric charges}

Equation (\ref{charge a}) then shows that electric charges must occur in
integer multiples of a basic charge $q=\hbar c$,\footnote{We can change to
more conventional units by rescaling charges and potentials according to
$\lambda q_{n}=q_{n}^{\prime}$ and $A_{a}/\lambda=A_{a}^{\prime}$ so that
$q_{n}A_{a}=q_{n}^{\prime}A_{a}^{\prime}~$. For conventional units such that
the basic charge is $q^{\prime}=e/3$ with $\alpha=e^{2}/\hbar c=1/137$ the
scaling factor is $\lambda=\frac{1}{3}(\alpha/\hbar c)^{1/2}$.}
\begin{equation}
q_{n}=\beta_{n}\hbar c~. \label{charge b}%
\end{equation}
Therefore, \emph{the condition for compatibility of the probabilistic and
linear structures that leads to full equivalence between ED and the QM of
charged particles is that charges be quantized.}

\section{Conclusions}

\paragraph*{On quantum geometry--}

The information geometry of the space of probabilities can be naturally
extended to the full e-phase space of probabilities and phases. This
reproduces the Riemannian, complex and symplectic structures characteristic of
QM. The ED that preserves these structures while recognizing the privileged
ontic role of position is the Schr\"{o}dinger equation.

\paragraph{On charge quantization and single-valuedness--}

There is a deep connection between the probabilistic structure of QM, its
linearity, the single-valuedness of wave functions, and the quantization of
charge. Full equivalence between ED and QM is achieved when we adopt
constraints that reflect the intimate relation between quantum phases and
gauge invariance, and also reflect the empirical fact that charges are
quantized. If, alternatively, we take linearity as the \textquotedblleft
empirical\textquotedblright\ input, then we deduce charge quantization.

\paragraph{Acknowledgments}

I would like to thank M. Abedi, D. Bartolomeo, C. Cafaro, N. Carrara, N.
Caticha, S. DiFranzo, A. Giffin, S. Ipek, D.T. Johnson, K. Knuth, S. Nawaz, M.
Reginatto, C. Rodr\'{\i}guez, and K. Vanslette, for many valuable discussions.

\end{document}